\documentclass[prb]{revtex4}

\usepackage{graphics}

\newcommand{\rb}{\mbox{\boldmath $r$}}
\newcommand{\Ob}{\mbox{\boldmath $0$}}

\begin{document}

\title{Scaling laws at the critical point}

\author{S.\ Davatolhagh}
\affiliation{Department of Physics, College of Sciences,
Shiraz University, Shiraz 71454, Iran}


\begin{abstract}
There are two independent critical exponents that describe the behavior of
systems near their critical point. However, at the critical point only the
exponent $\eta$, which describes the decay of the correlation
function, is usually discussed. We emphasize that there is a second
independent exponent $\eta'$ that describes
the decay of the fourth-order correlation function. The
exponent $\eta'$ is related to the exponents determining the behavior of
thermodynamic functions near criticality via a fluctuation-response
equation for the specific heat. We also discuss a scaling law for $\eta'$.

\end{abstract}

\maketitle

It is well known that the critical behavior of a system near its
continuous phase transition\cite{JT} can be completely described
by only two independent exponents.\cite{KH} This description can
be expressed by the scaling behavior of the singular part of the
free energy density
\begin{equation}
f_{\rm
sing} = |t|^{2-\alpha} f_\pm (h/|t|^{\Delta}), \qquad
(t\rightarrow 0^{\pm})
\label{1}
\end{equation}
where $t \equiv (T-T_c)/T_c$ is the reduced temperature, $T_c$ is
the critical point temperature, and $h$ is the external field that
is coupled to the order parameter. The scaling functions
$f_\pm(x)$ depend on $t$ and $h$ only in the combination
$x=h/|t|^{\Delta}$. [xx note change of notation xx] Because the
singular contributions to various thermodynamic quantities are
found from various derivatives of the free energy in
Eq.~(\ref{1}), there are only two independent exponents, say
$\alpha$ and $\Delta$, that characterize the critical behavior of
the system and thus specify its universality class. For example,
the dependence on $t$ in Eq.~(1) leads to a power-law singularity
for the specific heat, $C_{\rm sing} \sim -\partial^2 f/\partial
t^2$ when $h=0$:
\begin{equation}
C_{\rm sing} = |t|^{-\alpha} c_\pm (h/|t|^{\Delta})
= c_\pm(0)|t|^{-\alpha}, \qquad (h=0).
\end{equation}

At the critical point the thermodynamic derivatives are not useful
for characterizing the critical behavior as they usually take on
infinite values. [xx most of the derivatives are infinite. don't
understand this sentence xx] The only well known exponent that is
defined at the critical point is the anomalous dimension exponent
$\eta$, which describes the decay of the order parameter
correlation function:
\begin{equation}
G_{ m} (\rb) \equiv
\langle\delta m(\rb)\delta m(\Ob)\rangle = \langle m(\rb)
m(\Ob)\rangle - \langle m \rangle^2 \sim 1/r^{d-2+\eta},
\label{eq:3}
\end{equation}
where $\delta m(\rb) = m(\rb) - \langle m \rangle$ is the
order-parameter density fluctuation from the mean at point $\rb$.
The angular braces $\langle\cdots\rangle$ denote an ensemble
average, and $d$ is the spatial dimension. Note that $G_{\rm m}
(\rb)$ is a two-point correlation function. Although we will use a
notation appropriate to magnetic systems, the following discussion
also is applicable to other systems that have a continuous phase
transition, such as fluids at their critical point.

The search for a second independent exponent leads to the
fourth-order correlation function in terms of the order parameter
density with an exponent defined by:\cite{MK}
\begin{equation}
G_{\rm E}(\rb) \equiv \langle H(\rb)
H(\Ob)\rangle - \langle H \rangle^2 \sim 1/r^{d-2+\eta'},
\label{eq:4}
\end{equation}
where $H(\rb)$ is the local Hamiltonian or the
interaction energy density. Note that the local
interaction energy density $H(\rb)$ ($\sim
m(\rb)^2$) is quadratic in the order parameter density $m(\rb)$.

The exponents $\eta'$ and $\eta$ are related to the usual exponents via a
fluctuation-response equation for the order parameter susceptibility
\begin{equation}
\chi =
\beta\!\int d^d r G_{\rm m} (\rb) \sim \!\int^{\xi} \frac{d^d
r}{r^{d-2+\eta}} \sim\xi^{2-\eta}, \label{eq1}
\end{equation}
and the energy version of Eq.~(\ref{eq1}) which relates the
thermal susceptibility or the specific heat:
\begin{equation}
C/k = \beta^2\!\int d^d r G_{\rm E} (\rb) \sim
\!\int^{\xi} \frac{d^d r}{r^{d-2+\eta'}} \sim\xi^{2-\eta'}.
\label{eq2}
\end{equation}
The correlation length $\xi$ characterizes the
exponential decay of the correlations, $\beta=1/kT$, and $k$ is Boltzmann's
constant. Equation~(\ref{eq1}) yields the well-known Fisher scaling
identity:\cite{Fisher}
\begin{equation}
\gamma = (2-\eta)\nu, \label{eq3}
\end{equation}
where $\gamma$ is the linear response susceptibility exponent
defined by $\chi\sim |t|^{-\gamma}$, and $\nu$ is the correlation
length exponent, $\xi\sim |t|^{-\nu}$. Equation~(\ref{eq2}) yields an energy
version of Fisher scaling:\cite{MK,D05,Card}
\begin{equation}
\alpha = (2-\eta')\nu. \label{eq4}
\end{equation}

The correlation length exponent $\nu$ appearing in
Eqs.~(\ref{eq3}) and (\ref{eq4}) must be the same because of the
hyperscaling hypothesis, which states that close to criticality
the correlation length $\xi$ is the only relevant length scale,
and thus $\xi$ is solely responsible for the singular
contributions to the thermodynamic quantities. Another important
outcome of the hyperscaling hypothesis is the Josephson scaling
identity, $2-\alpha = d\nu$, which provides another relation
between $\alpha$ and $\nu$ and thus gives a constraint on
Eq.~(\ref{eq4}). If we substitute $\alpha = 2 - d\nu$ into
Eq.~(\ref{eq4}), we obtain another scaling law involving $\eta'$:
\begin{equation}
\eta' =
d+2-\frac{2}{\nu}. \label{eq6}
\end{equation}
We see that the critical
exponent $\eta'$ is determined by the
correlation length exponent $\nu$ and the spatial dimension $d$.

To verify the scaling equation (\ref{eq6}) and to
test its predictions, we first consider
the $d=2$ ferromagnetic Ising model in the absence of
an external field for which exact analytic solutions are
available.\cite{LO,2DIsing} In particular, we are interested in the
four-point correlation function of the form
\begin{equation}
w_4(r) = \langle S_1 S_2 S_r S_{r+1}\rangle - \langle S_1
S_2\rangle \langle S_r S_{r+1}\rangle
\end{equation}
which also can be interpreted as a two-point
energy correlation, with $S_1 S_2$ characterizing the energy of a
reference bond and $S_r S_{r+1}$ the energy of a bond
at a distance $r$ from the reference bond, as can be seen in Fig.~1.
The Ising spins take on the values,
$S_r = \pm 1$. The fourth-order correlation function
$w_4 (r)$, when summed over $r$ for all distinct pairs of bonds
is related to the specific heat in agreement with Eq.~(\ref{eq2}): $\sum_r w_4 (r) =
\partial\epsilon/\partial\beta - 1 + \epsilon^2$, where $\epsilon
= \langle S_1 S_2\rangle$ characterizes the mean energy per
bond.\cite{TB}

From Eq.~(\ref{eq6}) we see that $\eta'=2$ for the $d=2$ Ising
model, where we have used the exactly known correlation length
exponent $\nu=1$. We note that when $\eta' =2$ is substituted into
the Eq.~(\ref{eq2}), we find a logarithmic divergence for the
specific heat as we should. The asymptotic behavior of $w_4(r)$
has been calculated rigorously for the triangular and the square
Ising lattice and has been found to be identical for the two
lattices (apart from an amplitude factor of $1/2$) and has the
usual Ornstein-Zernick forms\cite{JS}
\begin{equation}
w_4(r)\sim \frac{e^{-2r/\xi}}{r^2}, \qquad (r\rightarrow\infty).
\label{eq8}
\end{equation}
We see that the predicted value $\eta' = 2$
agrees with results from exact analytic calculations.

For comparison the critical exponents relevant to our discussion
for the $d=2$ and $d=3$ Ising models are summarized in
Table~\ref{table1}. For $d=3$ Eq.~(\ref{eq6}) predicts $\eta' =
1.822\,(3)$, where we have used the numerical estimate, $\nu =
0.629\,(1)$.\cite{3DIsing} A direct measurement of the ratio
$\alpha/\nu = 2-\eta'$ can be obtained by a finite size scaling
analysis\cite{FSS} of the specific heat obtained by Monte Carlo
simulations.\cite{LB} The slope of a log-log plot of $C$ versus
$L$ provides an estimate for the ratio $\alpha/\nu$: $C \sim
t^{-\alpha} \sim L^{\alpha/\nu}$. A reliable estimate of this
ratio for $d=3$ is $\alpha/\nu = 0.178\,(5)$.\cite{3DIsing} Thus
we have $\eta' =2-\alpha/\nu = 1.822\,(5)$, which is in good
agreement with the value predicted by Eq.~(\ref{eq6}).

In summary, the exponent $\eta'$ of the four-point correlation
function provides a second independent exponent at the critical
point. Thus \{$\eta'$, $\eta$\} form an independent exponent pair.
The discussion presents a convenient framework for introducing the
concept of fourth-order correlation functions, which are important
in the study of viscous liquids\cite{D05} and glasses.\cite{DFP}

\begin{table}[h]
\caption{Comparison of various critical point exponents
for the $d=2$ and $d=3$ Ising models.
\label{table1}}
\begin{ruledtabular}
\begin{tabular}{lccccc}
System & $\alpha$& $\nu$& $\eta$& $\eta'$\footnotemark[1]&
$\eta'$\footnotemark[2]\\
\colrule
2D Ising model (exact)& 0& 1& 1/4& 2& 2 \\
3D Ising model (approx)& 0.112(3)& 0.629\,(1)& 0.037\,(1)& 1.822\,(3)& 1.822\,(5) \\
\end{tabular}
\end{ruledtabular}
\footnotetext[1]{From Eq.~(\ref{eq6})} \footnotetext[2]{From
analytical\cite{JS}/numerical\cite{3DIsing} calculations}
\end{table}

\begin{figure}[h]
\includegraphics{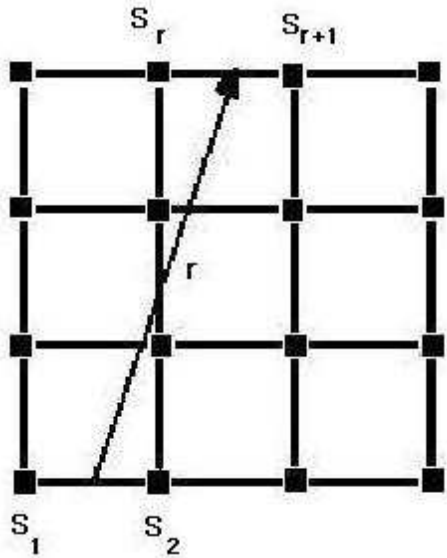}
\caption{Schematic of four-point spin correlations in the $d=2$ Ising model.
The line segments represent bonds whose energies are determined by the
product of nearest-neighbor spins at each end.}
\end{figure}

\end{document}